# IRON-FREE DETECTORS FOR FUTURE LINEAR COLLIDERS

Alexander Mikhailichenko, Cornell University, Ithaca, NY

*Abstract.* We continue consideration of Iron-free magnetic systems for possible application in detectors for High Energy Collider. In particular we suggest a new type of magnetic system with the multiple flux-return solenoids. This system allows reaching higher field level at IP compared with traditional ones.

## INTRODUCTION

After discovery of Higgs-like particle, the Standard Model (SM) will serve for many decades as a basis for description of Nature. However the post SM word could be expected as much more interesting enterprise. So it is natural now days to develop techniques and technologies which will be helpful in discoveries of these post-SM visions. So the high energy colliders will be on agenda of physical community for long years ahead. Every next step in energy of colliders inevitably raises the question about best and adequate technology for appropriate detector, which accompanies the linear collider.

Proposal of 4[th] Concept for ILC [1], [2] demonstrated a step to practical implementation of Iron-free detector concept into high energy physics. With its 3.5 *T* field at IP it was able to deliver unique capabilities for particles identification. We believe that namely Iron-free detectors have a bright future in this post-SM discoveries accompanied by new acceleration technologies.

Doubts about necessity of Iron yoke in detector magnets rose a long ago, however [3], [4], [16]. With introduction of dual readout techniques [5]-[9] the idea of Iron free detector come to its practical stage with 4[th] Concept.

It is interesting to mention here, that the Iron-free solenoidal systems are requested nowadays by commercial applications in MRI technique [10]. Impressive 11.7 *T* solenoidal system has no Iron yoke at all (INUMAC project). Desire to exclude the Iron yoke is natural: the Iron saturated at ~2*T* and further rise of field is going with effective magnetic permeability $\mu \sim 1$, although saturated Iron still delivering this 2*T* magnetization.

Among many innovations, including Iron-free magnet, and dual readout system of identification, the innovation which, we believe, will be useful in a future -is associated with the cluster counting (*CluCou*) as a tracking system [11]. This methodic allows delivering a resolution of wire chamber much smaller, than the transverse dimensions (diameter) of the wire chamber cell itself. In addition, usage of CluCou instead of TPC[1] allows relaxed restrictions on homogeneity of magnetic field at tracking region of detector.

Longitudinal field at Interaction Point (IP) of any collider has a high level required by proper identification of momenta of the secondary particles generated at IP. Longitudinal magnetic field well fits into axial symmetry of final focusing system of colliding beams. The magnetic field value together with the size of tracking system, defined by required momentum resolution, which is $\Delta p / p \sim p\sigma_s /(B_0 D^2)$, where $B_0$ stands for the axial field in a central solenoid, $D$ is its diameter, $\sigma_s$ is a spatial resolution of tracking system. Typically, magnetic field created with the help of superconducting solenoid with induction of 4*T* (ILD)-5*T* (SiD). Magnet yoke of detectors for colliders have tens of thousand tons of

---

[1] Time Projection Chamber [12]

Iron to re-direct the magnetic field flux from the one end of solenoid to the opposite one. From the other hand it is known, that the magnetic field value outside of the (long) solenoid is zero. Solenoids used (or suggested for use) have some remaining field outside, depending on the length/diameter ratio. In practice, the iron adds ~20% of the field value in a realistic geometry only. With invention of dual-readout calorimeters which are able to determine the type of particle, identification of muons, carried usually with the help of chambers implemented into back leg of yoke iron, is now transferred to the calorimeter itself.

In this communication we represent the basic principles put in grounds of iron-free detector with multiple flux-returns system. In such detector the magnetic flux is closed with the help of many additional solenoids. Stray field outside detector has minimal level with implementation of end wall-of-coils. With elimination of iron yoke the detector becomes a lightweight unit and all elements of it easy accessible for further modifications (for different energy of colliding beams for example). Namely engineering realization and some technologies associated with such detector, suggested for ILC (4$^{th}$ Concept) described in more detail.

We are projecting parameters of such detector for usage with a multi TeV-scale colliding beams which inevitably will appear in a future for investigations in a post-Standard Model of the Universe.

## OVERVIEW

The steel yoke of any contemporary detector for High-Energy physics impresses everyone who had a chance to see it closely. Structurally this (few-tenth of kiloton in case of ILC) detector consists few main elements such as:
1) Pixel vertex detector for high-precision identification of vertex;
2) Tracking system immersed into magnetic field for 3D restoration of tracks;
3) Calorimeter(s) for the energy measurement of hadrons, jets, electrons, photons, missing momentum, and the tagging of muons and other particles;
4) Iron yoke with incorporated muon system.

Typical detector cross section represented in Fig.1.

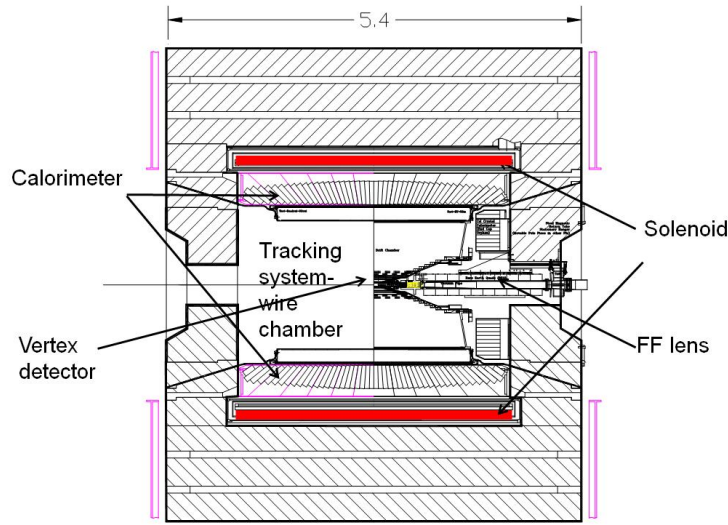

**Figure 1.** Typical mid-size detector (CLEO [13]; operated for ~5$GeV$ e$^+$e$^-$ beams). Dimension is given in meters. Iron yoke is hatched. If permeability of Iron put to a one, the field at the center will be 25% lower for this particular geometry.

The main role of this magnet yoke is in service as a duct for the flux return for main (central) solenoid. If the magnetic permeability of yoke changed to a one (Air), the field inside a solenoid will drop about 25% only for a typical detector from Fig.1. Mostly this change impacts the field homogeneity. This drop associated with the finite ratio of the length of main solenoid to its diameter. It is well known that there is no significant field outside of long solenoid. Field outside has strictly zero value for (infinitely) long one. Also, the field is homogenous inside the (long) solenoid. So bigger the Length/diameter ratio-lesser the drop is. Homogeneity could be restored by adding current caring coils at the ends of central solenoid (Helmholtz coils), however.

Having good field homogeneity in a region, where the tracking system is located, required for the tracks identification is easier. With a Cluster Counting *CluCou* technology [11], see below, the homogeneity required could be less than with the TPC. The productivity of contemporary processors dedicated to this job, allow corrections effects of field inhomogeneity to be done in a real time.

In addition, detector has typically (superconducting) final quads located inside the magnet field of detector and theirs field have significant value in a region where the wire chamber located. This makes trajectory analysis more complicated also. Detector physicists are prepared for this and are ready to make all necessary corrections, (what indicates a potential for further developments).

Thinking ahead, with some novel accelerator techniques, see for example [18], one should foresee a possibilities for detectors, having multi-$TeV$ colliding particles. These detectors will require as high field in central region as possible with maximal possible diameter of central solenoid. One can count on implementation of 10-20 $T$ fields in central SC solenoid in a future. Meanwhile the iron becomes deeply saturated at the field level ~$2T$, so the magnetic yoke of a traditional detector will manifest saturation even for the field level ~$3T$.

We believe that the concept of Iron-free system is an inevitable way to go for high-energy detectors in a future.

## THE CONCEPT OF IRON-FREE DETECTOR

The yoke is an element of the magnet circuit only, so anyone can consider a review for its elimination. For realistic diameter/length ratio homogeneity of field in a central region will drop, naturally with elimination of Iron. However with additional ampere-turns at the end region of superconducting solenoid (Helmholtz-type) the field can be made homogenous again to any level required. Additional heat and electricity losses are tolerable. These additional turns can be located, naturally, inside the same cryostat[2]. Few possibilities become open for the Iron free detector design.

A family of Iron-free detectors is represented in Fig.2. It begins from just a single, solenoid, a). This single-solenoid system is inexpensive, compact, but it generates significant stray-field in outer space. This stray field requires attention, but could be screened by relatively thin sheets of iron. Dual solenoid system b) is much better in this aspect. One minus of dual solenoidal system is that the field of outer solenoid, having opposite direction to the main solenoid reduces the field in a central region (about 1.6 $T$ for the 4[th]). Next member of this family is a triple-solenoidal system -c). Here two outer solenoids have opposite polarities of currents, so there is no reduction of field at IP. The field between the inner (first) and the second solenoid is about zero, i.e. it is like a free space. Minus of this system is that it requires additional solenoid, although the field generated by this outer solenoid is few times weaker, than the inner one. The next member of the family is a one with multiple flux-return solenoids-d). This type requires fabrication of many solenoids, but as the diameter of solenoids is small, these ones could be fabricated with much less effort, than the additional solenoid in b) and c). Finally, the last magnetic system-e), represents the multiple-return solenoid system with segmented solenoids for better coverage of volume by magnetic field.

---

[2] Some distribution of local current density along solenoid could be appointed also.

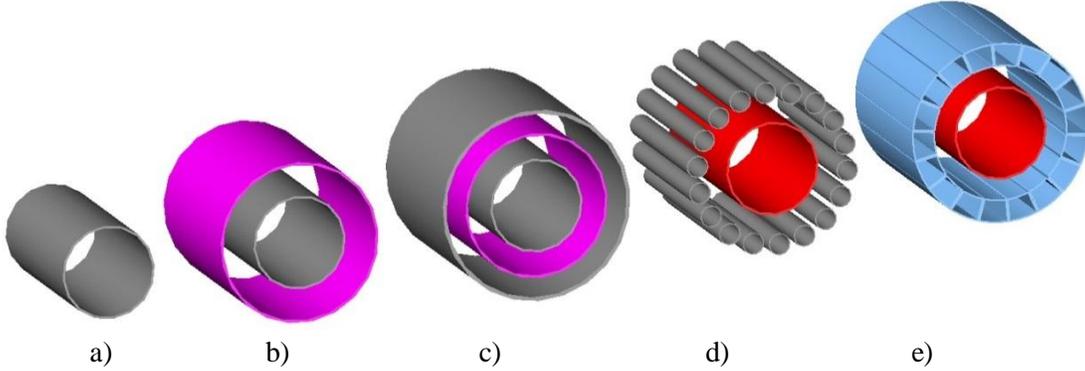

a) b) c) d) e)

**Figure 2.** A family of Iron-free detectors: a)-single solenoid, b)-dual solenoids, c)-triple solenoids, d)- many return-flux solenoids. e)-many return-flux solenoids with sectorial shape. Each system of solenoids surrounded by the end-cap-wall of coils (which are not shown here, see Fig.4).

Basic parameters important for the field calculations in a multiple-solenoidal system could be seen in Fig.3.

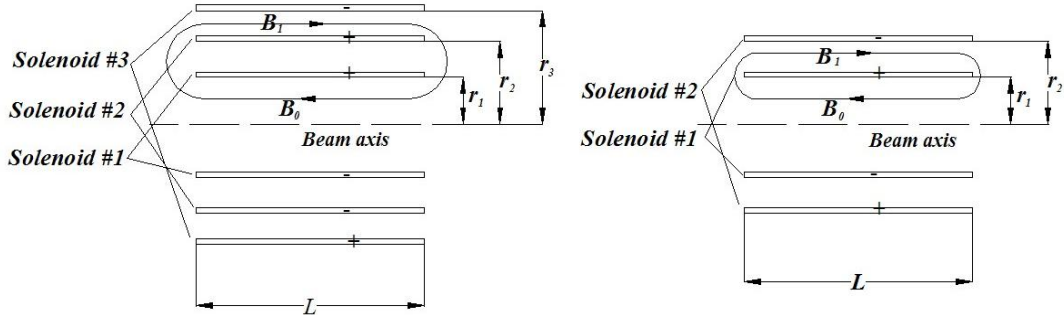

**Figure 3.** Geometry of three–coil system, left. At the right there is represented the situation when two coils from the left figure merged together ($r_1=r_2$). Signs "+" and "–" indicate direction of solenoidal current circulating in the coil.

Magnetic field and the current in each solenoid can be found from a simple condition

$$B_0 \times r_1^2 = B_1 \times (r_3^2 - r_2^2) , \qquad (1)$$

which is just a reflection of conservation of the flux. When two coils merge together, the last formula simplified to the following

$$B_0 \times r_1^2 = B_1 \times (r_2^2 - r_1^2) , \qquad (2)$$

Magnetic field $B \cong NI/L$, where $NI$ stands for the total current running in the coil, $L$ is the (effective) length of solenoid. So the volume between coils at $r_1$ and $r_2$ (solenoid 2 and 1) can be made practically free from magnetic field. The last circumstance might be useful in some cases.

Let us estimate the fields ratio for typical values which are $r_1 \cong 2.5$m, $L \cong 5$ m, $B_0 \cong 5$ T. So if $r_2 \cong 4$m (1.5 m radial space between inner solenoid and the next one), $r_3 \cong 5$m, then in first case (thee coils), magnetic field value in return space between solenoid 3 and 2 comes to

$$B_1 = B_0 \frac{r_1^2}{r_3^2 - r_2^2} \cong 5 \frac{2.5^2}{5^2 - 4^2} \cong 3.5 \, T \tag{3}$$

and in the second case (two coils) magnetic field goes to

$$B_1 = B_0 \frac{r_1^2}{r_2^2 - r_1^2} \cong 5 \frac{2.5^2}{4^2 - 2.5^2} \cong 3.2 \, T \tag{4}$$

One can easily scale these figures to any appropriate radii. One might consider the placement of two outer solenoids practically at the outer housing of detector.

The field outside of solenoid drops rapidly, however. Basically it drops as a third power of the distance $R$ to the point of observation,

$$\vec{H} = \frac{3\vec{n} \cdot (\vec{n} \cdot \vec{M}) - \vec{M}}{R^3}, \tag{5}$$

where $\vec{n}$ is an unit vector in direction of $R$, and $\vec{M}$ is the magnetic moment of solenoid,

$$\vec{M} = \frac{1}{2c} \int (\vec{j} \times \vec{r}) dV = \frac{\pi r^2 J}{2c}, \tag{6}$$

$J$ is total current, $\vec{j}$ is a current density, $r$ is the radius of solenoid. Even at the distance of ~1-2 meters the fields naturally drops to ~$0.5 kG$, where the local iron shields can be implemented easily, if necessary. Some local shielding far from the solenoid ends can be implemented easily.

We would like to remind that the Iron itself might cost $35M easily; one can refer to this number in publications at ILC web-site. The cost of Iron-free detector with SC coils expected to be lower, than traditional detector with Iron. At least one SC coil is present in any detector anyway (main solenoid), so the cost of other two must be compared with the cost of iron, its tooling, transportation, and installation.

Mostly impressive advantage of Iron-free detector is a functional flexibility, easy commissioning in addition to lowered cost. The last allows fabrication of two (or even more) detectors for experiments. We called this concept a *modular detector*.

Field inside inner and outer solenoids can be made homogeneous to the level required by adding the coils at the end of each solenoid (Helmholtz-type coils). Optimization of such system takes very short time with appropriate codes (MERMAID, FlexPDE). Magnetic mapping allow proper reconstruction of trajectory practically with any field distribution, however.

We will describe technologies accepted in 4[th] Concept detector in a view of implementation of these technologies in the Iron-free detector with multiple flux-return solenoids.

## COMPONENTS AND TECHNOLOGIES

Let us underline some basic principles and technological solutions affecting design of magnetic system of future detectors. All these are implemented in the 4[th] Concept detector.

### *Dual readout calorimeter*

Dual readout techniques [5]-[8] deal with the time structure analyses of signal from the crystals. Typically it is a fast Cherenkov light output and slower scintillation signal. So by measurements of signal in two different time gates allows distinguishing between the types of particle.

Typically, for the *gamma and lepton calorimetry*, the crystals of BGO ($Bi_4Ge_3O_{12}$) – Bithmuth Germanium Oxide could be used. This inorganic chemical compound is no hygroscopic and easy machinable.

Time structure of signals from BGO is represented in Fig. 4.

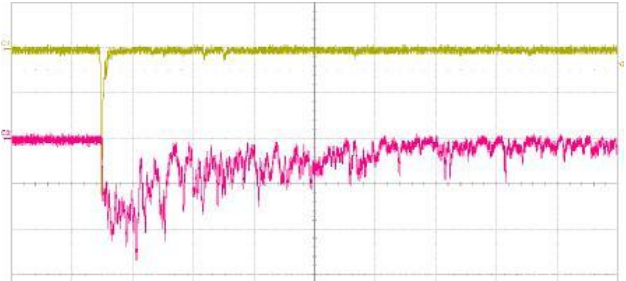

**Figure 4.** Time structure of Cherenkov signal (upper curve) and a scintillation one (the lower curve) [1], [5].

Other dual readout system is a *hadronic calorimeter* with two (or more) types of fibers having different properties for registration of Cherenkov light and scintillation. Utilization of (optical) filters allows better identification of Cherenkov light and scintillation one as they have different spectrum.

These fibers, having diameter~1*mm* and are ~1.5*m*-long, allocated in long cylindrical holes in angularly and longitudinally segmented towers assembled in a core inside main solenoid, with segments fabricated from hi-Z materials (Lead, Tungsten[3]). The scintillating fibers are grouped in Cherenkov and scintillating bunches delivering light to respecting photomultiplier.

The dual-readout crystal lepton calorimeter located in front of the fiber (hadronic) one.

*Cluster counting (CluCou)*

*CluCou* is a procedure for measurements and recording the drift times of all electron clusters generated by a particle on its way inside the drift tube or wire chamber [3], [9].

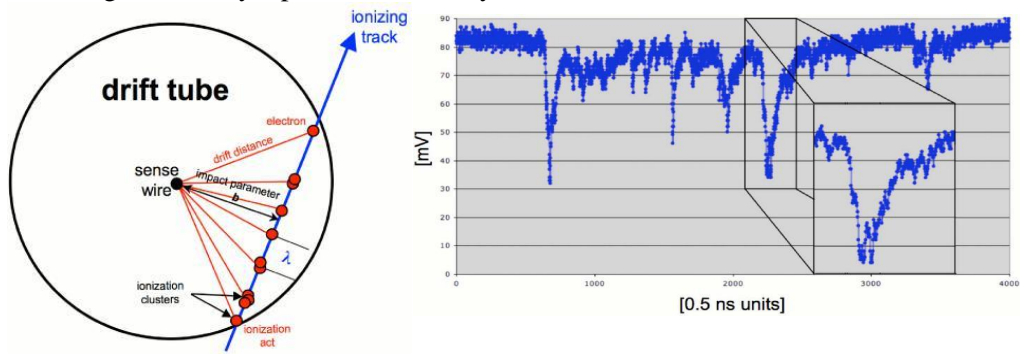

**Figure 5.** The principle of CluCou. At the left: geometry of drift tube with ionizing track. At the right: the time structure of a signal from drift tube [1], [11].

---

[3] The Lead is preferable, as it allows easy machining, even casting. Also, the Tungsten alloy contains some fraction of magnetic additions, so it requires a special control.

Typical gas mixture contains Helium (90%) with Iso-Butane (10%) $HeC_4H_{10}$. Wires are made from Carbon composite for lowering the amount of scattering substance. This method allows reaching a spacial resolution much higher than the wire granulation (tube diameter). So the CluCou will serve well as a tracking system in a future detector.

*Vertex detector*

The vertex detector is a multi Giga-pixel chamber with cylinders and disks [1]. With pixels of ~20 $\mu m$, spatial resolution could reach ~5$\mu m$. For a pixel size of 20 $\mu m$ with a dead area of 10 $\mu m$ along the perimeter of the sensors, the total number of channels comes to $4.3 \times 10^9$. In a future these pixel dimensions will be lowered as the technology progresses.

*Machine-detector interface (MDI)*

Requirements for MDI underlined in [14], [15]. One general requirement is that the Linear Collider should serve for at least two different detectors, although there is no requirement that they should do this simultaneously (Push-Pull concept). We think that this concept will be useful for detectors working with multi-*TeV* beams. Obviously, the off-beam line detector should be shifted in transverse direction to a garage position, located 15 *m* from the IP. The radiation and magnetic environment, suitable for people access to the off-beam line detector during beam collision in the other one, are guaranteed to be safe.

We anticipate that with development of more compact and, hence, less expensive Final Focus hardware (see [18]), these two detectors can be served by beams at the same time on the basis of fast Switch Yard system. In this case all the movement apparatus could be excluded, as the detectors stay in place.

## 4$^{TH}$ CONCEPT: DUAL SOLENOID SYSTEM

Let us consider the 4$^{th}$ Concept detector for ILC as example of dual solenoidal system. Detector developed for ILC by 4$^{th}$ concept team [1],[2] is a bright representative of dual-solenoid family, b). This detector represented in Fig 6.

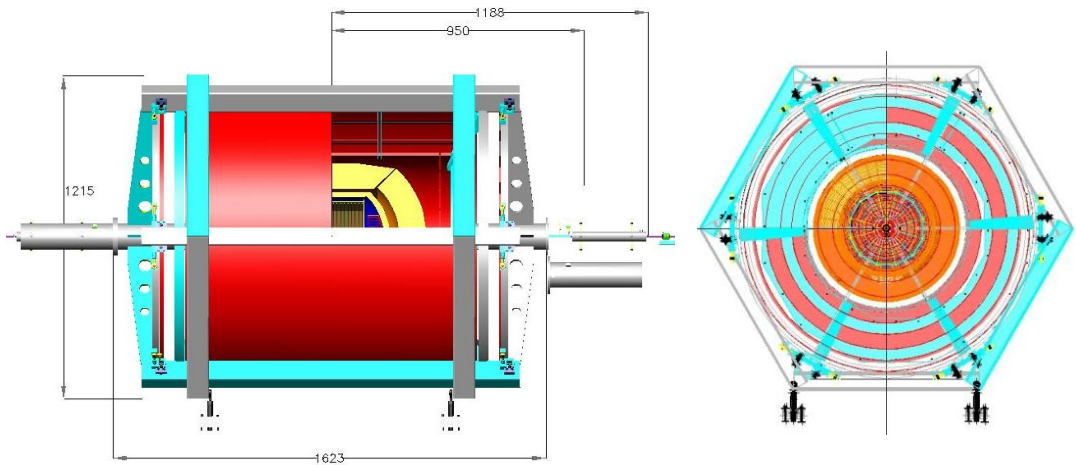

**Figure 6.** The 4$^{th}$ Concept detector suggested for ILC [1]. Dimensions are given in *cm*.

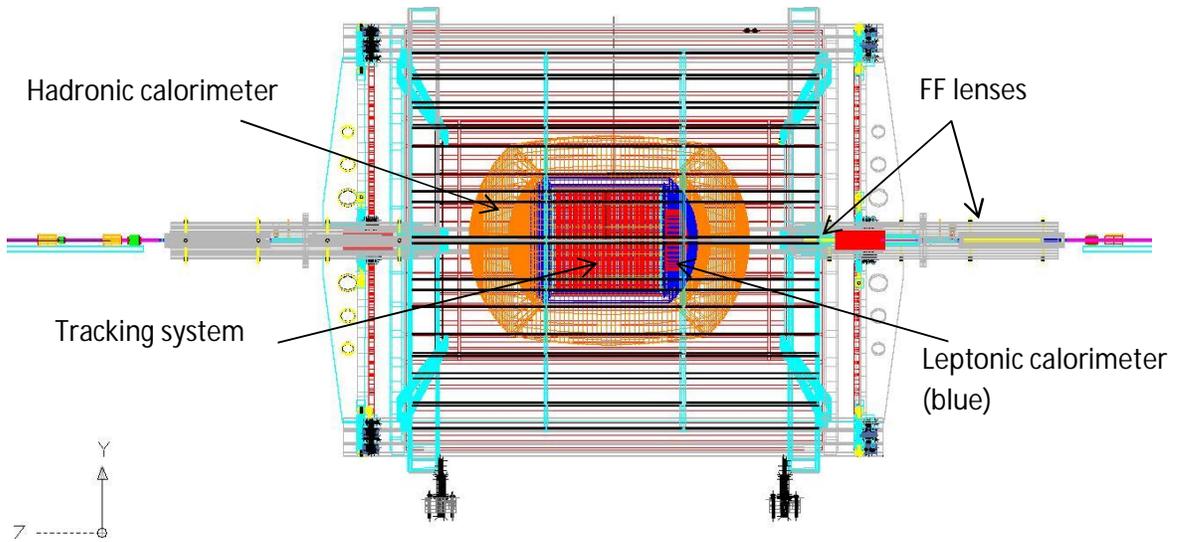

**Figure 7.** Transparent side view.

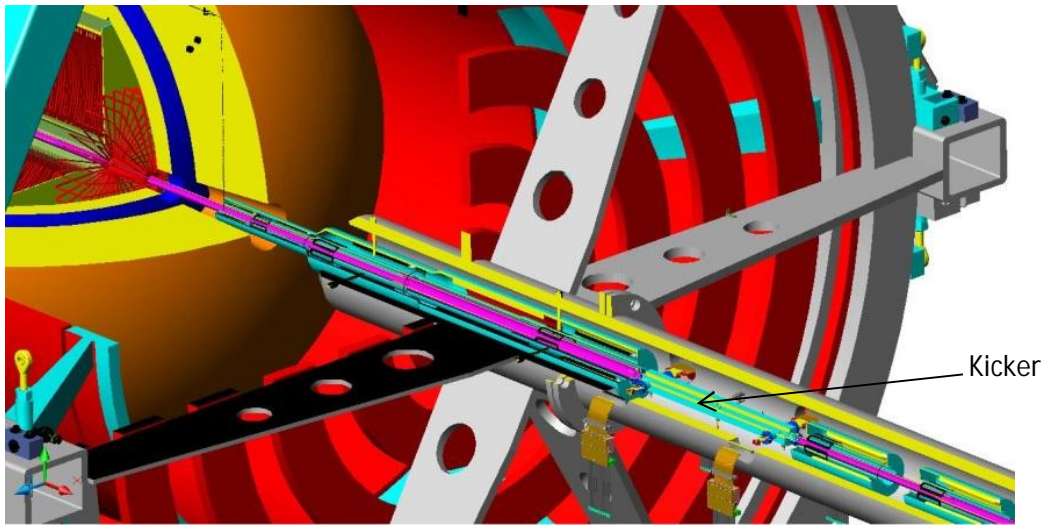

**Figure 8**. Isometric view. Final doublet with sextupoles, and the kicker (for head-on-collisions).

Main components of detector are represented in Fig. 9. We proposed a modular design, so any of these components could be easily installed into (and removed from) the frame. Frame itself could be split in two halves, so removal of solenoids and calorimeters could be done without movement of lower part of frame. Active feedback system for positioning elements of beam-focusing optics prevent disturbance which might be introduced by vibrations of frame.

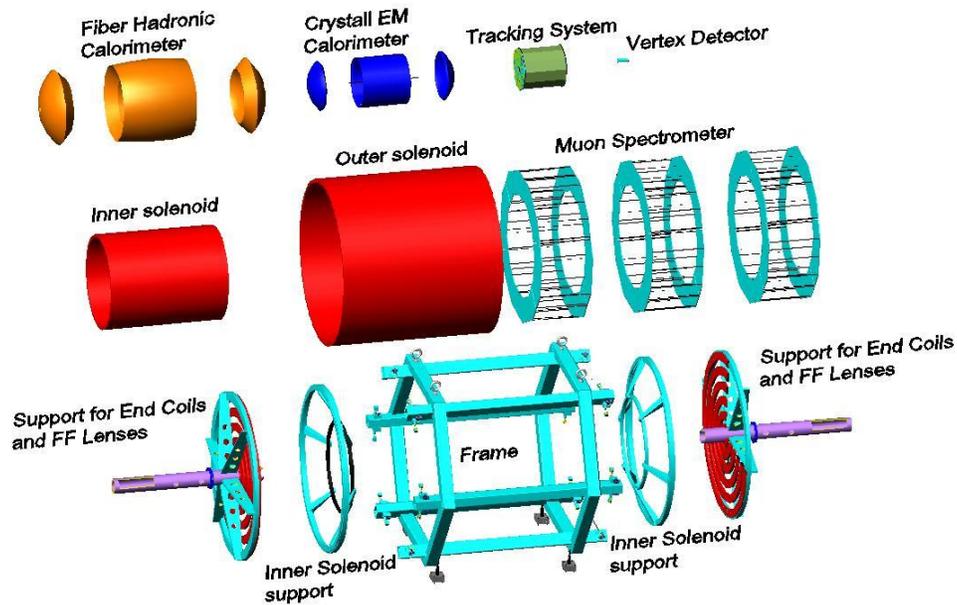

**Figure 9.** Main components of 4th Detector [1,2]. This structure could be recommended for a future detector.

*Magnetic field in 4th detector.*

Calculations of magnetic field were carried with help of MERMAID and FlexPDE codes. Dimensions of coils are represented in Fig.10.

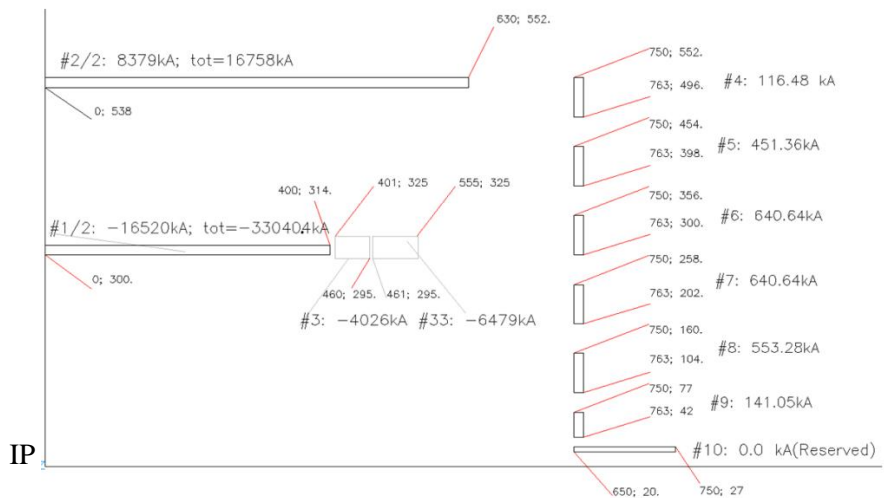

**Figure 10.** Locations and values of currents in 4th detector; ¼ of total cross section. IP located at lover left point of coordinate system. Numbers represent coordinates of points in *cm*, calculated from IP (one of variants).

The total stored energy in a magnetic field ~2.77 *GJ*. Namely this energy should be evacuated if quench occurred.

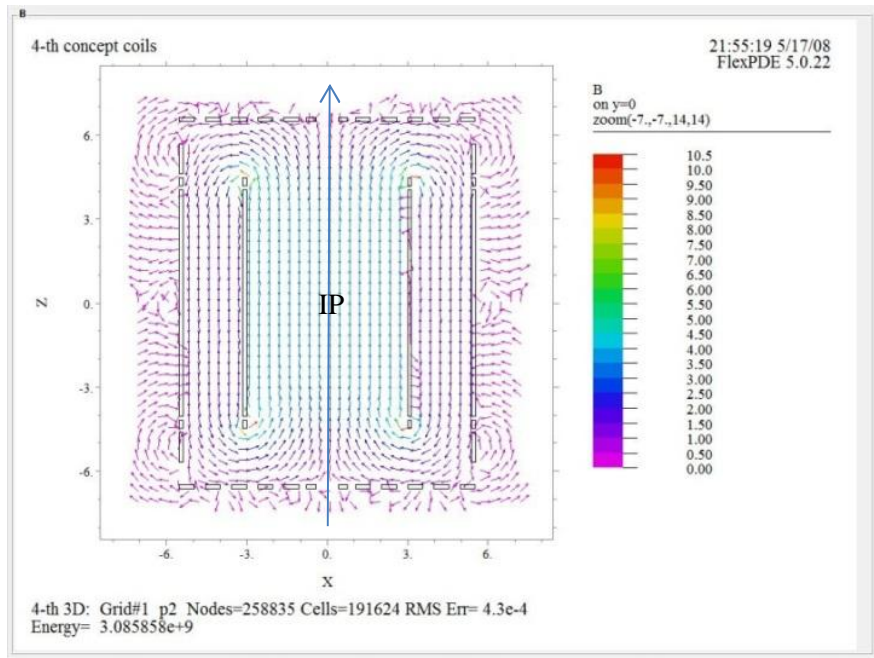

**Figure 11**. Vectors of magnetic field; full cross section. Arrow corresponds to the beam axis line.

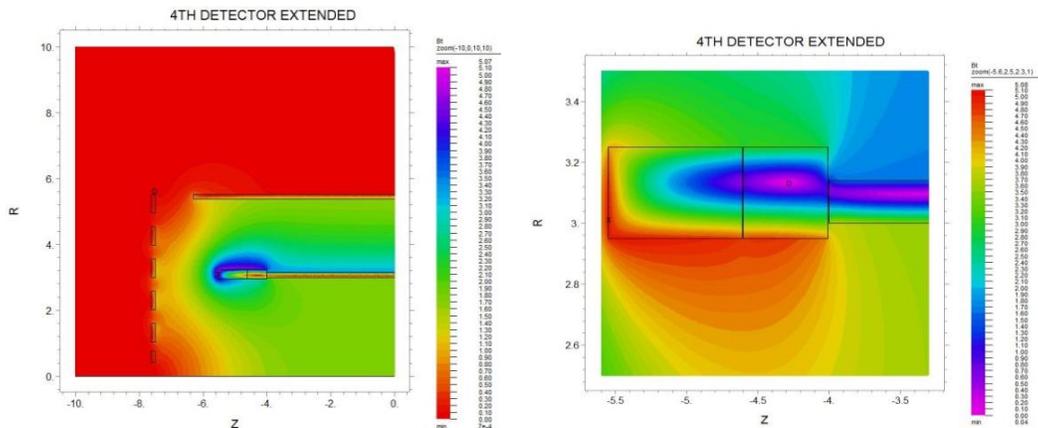

**Figure 12**. Contour plot of magnetic field module. Each color gradation corresponds to a $\Delta B / B \cong 6 \cdot 10^{-3}$ value in a field variation.

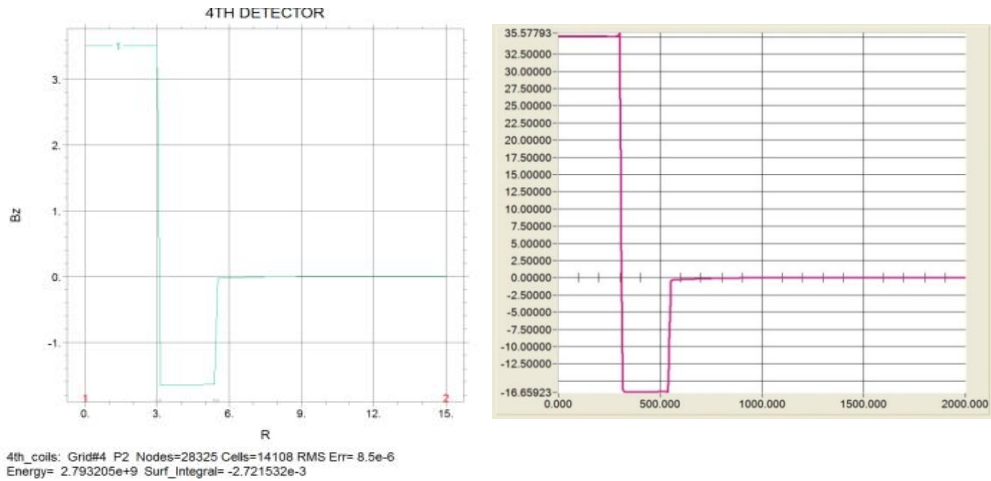

**Figure 13**. At the left: Radial distribution of longitudinal component of the field in a median plain, FlexPDE; Field measuresd in *Tesla*. At the right: The same distribution calculated with MERMAID; Field measuresd in *kGauss*.

All side coils are *room-temperature* ones; they have ~same current density; water cooled. Current density in coils (from the smallest radius to the biggest): 1; 8; 4.2; 3.3; 3.7; 1.7 $A/mm^2$, corresponding longitudinal forces are: 1.75; 102; 131; 135; 111; estimated weight~10 tons.

Field outside detector can be zeroed to any level by proper current distribution; Coils can be fixed easily at the end plates. These plates have reinforcement ribs helping in withstanding against magnetic pressure. Deformations of these plates are tolerable, see Fig.14 and could be corrected by active feedback systems.

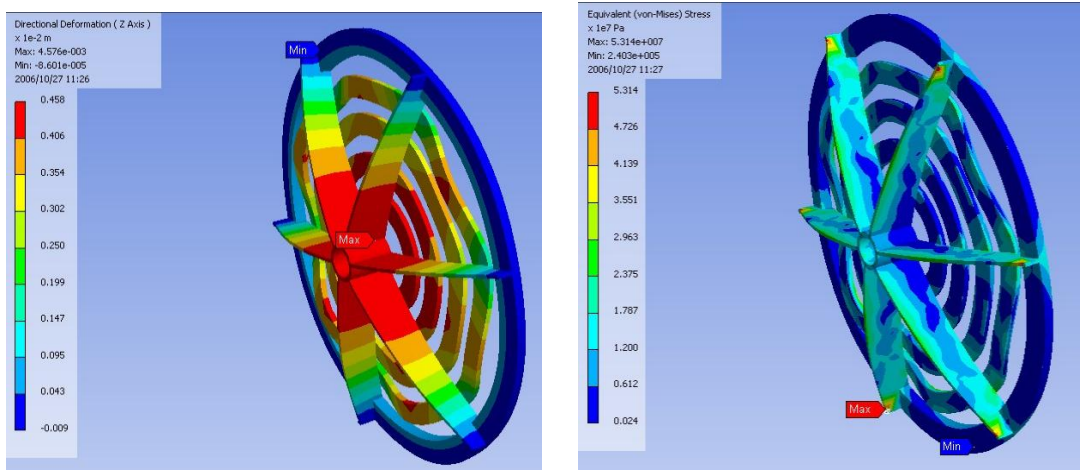

**Figure 14.** Deformation of frame with end coils [14]. Maximal deformation is 4.57 *mm*, maximal stress ~5x$10^7$ *Pa*.

Field elevation along axis is represented in Fig.15. One can see that the fringe field is very small, and traces of residual field could be easily captured by yoke of magnetic lenses or by additional wrappings by mu-metal, if necessary.

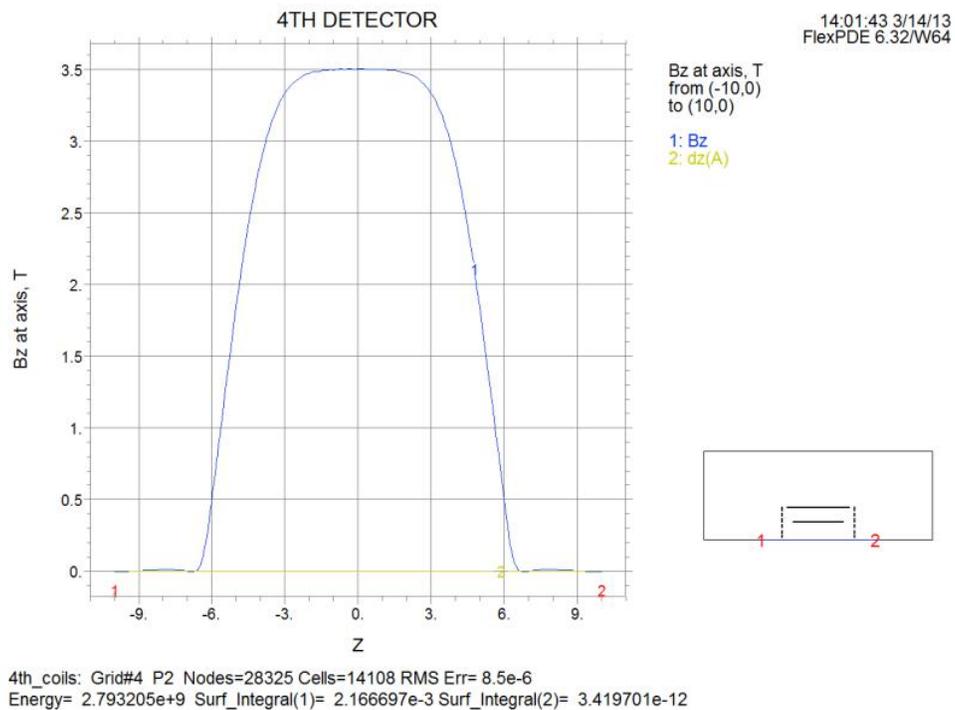

**Figure 15.** Longitudinal field distribution on axis. *z*-dimension is in meters, field in *Tesla*.

Space between solenoids used for muon spectrometry. Magnetic field level there is ~1.6 $T$[4]. This space filled with many tubes filled with a mixture of Helium and Iso-Butane He +$C_4H_{10}$ (90%+10%). Central wire of each tube for muon spectroscopy could be made from plated W. The number of tubes between solenoids comes to ~31500 tubes. The end caps contain 8640 tubes, Fig. 16 [1], [2].

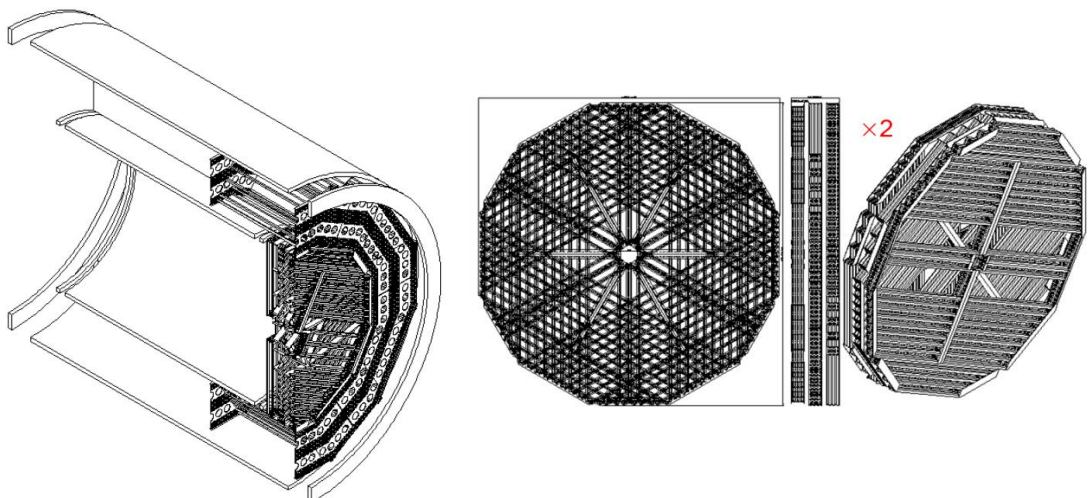

**Figure 16.** System of drift chambers between two solenoids (at the left). End caps magnified (at the right) [2].

---

[4] Namely this value is subtracted from the main solenoid.

## *Basic principles of 4$^{th}$, affecting* **MDI**
- Final Focus Beam-optical system is incorporated in Detector.
- Iron is omitted as it adds ~20% to the field value only (field outside of long solenoid is zero). Homogeneity restored by adding currents at the ends of main solenoid.
- Second solenoid closes the flux (minimal configuration).
- Muons can be identified with Dual (Triple) readout calorimeter scheme in a more elegant way

Usage of dual solenoidal system plus end wall current system allows:
1) Strict confinement of magnetic field inside limited region;
2) Spectroscopy of muons in magnetic field between solenoids;
3) Incorporate FF optics in mostly natural way;
4) Modular design which helps in modifications and re-installations;
5) Lightweight detector having flexible functionality and remarkable accuracy;
6) Easiest incorporation of laser optical system for gamma-gamma collisions.

Other specification reflected in [1] is that the superconducting final doublets of ILC, consisting on QF1 and QD0 Quadrupoles (and associated Sextupoles SD0 and SF1) are grouped into two independent cryostats. The cryostat with defocusing quadrupole QD0 penetrates almost entirely into the detector. The QD0 cryostat is a specific for the detector design and moves together with detector during push-pull operation, while the QF1 cryostat is a common one and rests in the tunnel, according to the current strategy of ILC BDR.
Meanwhile requirements for misalignment of these final Quadrupoles are severe.

## *Stability requirements for the lenses of final doublet*.

Final lenses (QF1 and QD0 in ILC project, see [1]) located at both sides of detector provide each-side beam focus at IP in both transverse directions – $x$ and $y$. If however, the quadrupole lens at one side is shifted transversely from its position, the kick for such displacement can be calculated as

$$\alpha = x' = \frac{e\Delta x \cdot \int G(s)ds}{mc^2\gamma} \cong \frac{\Delta x \cdot G \cdot l}{(HR)} \quad , \quad (7)$$

where $(HR)[Gs \cdot cm] = E[eV]/300$ is so called magnet rigidity of the high energy beam, $l$ stands for effective length of the lens, $G(s)$ describes its longitudinal field distribution with maximal gradient $G$ at the center. For $300 GeV$ beam magnetic rigidity comes to $(HR) \cong 10^9 [G \cdot cm] \equiv 10^6 [kG \cdot cm] \equiv 10^3 [T \cdot m]$, $\gamma \cong 6 \cdot 10^5$.

Propagation of kick $x'(s_0) = \alpha$ from its origin at the lens location $s_0$ to the IP located at $s_1$ counted from the lens's center, described by sin-like trajectory $S(s,s_0)$ having starting point at the lens location $s_0$

$$x(s_1) = x'_0(s_0) S(s_1, s_0) = \alpha \cdot S(s_1, s_0), \quad (8)$$

where $S(s_0, s_0) \equiv S(s_0) = 1$, $\alpha$ is a kick angle; with similar equation for the other transverse coordinate $y$ if kick happen in other direction too. By introduction of usual envelope function and the phase change as

$$\Delta\Phi \equiv \Delta\Phi_x(s_1, s_0) = \int_{s_0}^{s_1} ds/\beta_x(s), \quad (9)$$

displacement and the slope of the beam centroid at the IP come to

$$x(s_1) = \alpha\sqrt{\beta_x(s_1)\beta_x(s_0)}Sin(\Delta\Phi), \quad x'(s_1) = \alpha\sqrt{\frac{\beta_x(s_0)}{\beta_x(s_1)}}\left[Cos(\Delta\Phi) + \tfrac{1}{2}\beta'_x(s_1)Sin(\Delta\Phi)\right] \quad (10)$$

where $\beta_x(s_1)$, $\beta_x(s_0)$ stand for envelope functions values at the IP and at the lens respectively (for other coordinate, y, the functions are $\beta_y(s_1)$, $\beta_y(s_0)$). As the IP is the focusing point for this lens, then $Sin(\Delta\Phi) \cong 1$ as the betatron phase changes to $\Delta\Phi \cong \pi/2$ during transformation to IP.

If displacement is bigger, than the transverse beam size of incoming bunch (which is between 3.5–9.9 *nm*, according to BDR ILC), beams do not collide, so the requirement for the displacement at IP comes to

$$\sqrt{\frac{\gamma\varepsilon_x \cdot \beta_x(s_1)}{\gamma}} > \alpha\sqrt{\beta_x(s_1)\beta_x(s_0)}, \quad (11)$$

where $\gamma\varepsilon_{x,y}$ stand for invariant emittance for appropriate coordinate (left side is just beam size at IP). So the restriction for the kick and displacement come to

$$\alpha = \frac{\Delta x Gl}{(HR)} < \sqrt{\frac{\gamma\varepsilon_x}{\gamma\beta_x(s_0)}} \; ; \quad \Delta x < \frac{(HR)}{Gl}\sqrt{\frac{\gamma\varepsilon_x}{\gamma\beta_x(s_0)}} = \frac{1}{kl}\Delta x', \quad (12)$$

$\Delta x' = \sqrt{\frac{\gamma\varepsilon}{\gamma\beta(s_0)}}$ –is the angular spread at the location of lens, $k = \frac{G}{(HR)}$ –is the lens parameter, and the similar equations for *y*- coordinate. One can see that this restriction is not depending on beta-function value at IP.

Normalized emittance of ILC beam is $\gamma\varepsilon_x \cong 10^{-5} m \cdot rad$, $\gamma\varepsilon_y \cong 4 \cdot 10^{-8} m \cdot rad$, so the vertical jitter emerges as the mostly serous. Let us estimate the tolerances for QF1 as if it is based at the tunnel site and its jitter is not correlated with the location of other lenses. For gradient in lens $G \cong 10 kG \cdot cm$, effective length of lens *l*=200*cm*, $\beta(s_0) \cong 10^4 m$, for 300-*GeV* beam energy, the vertical jitter (coordinate *y*) limited to

$$\Delta y < \frac{10^6[kG \cdot cm]}{10[kG/cm] \cdot 200[cm]}\sqrt{\frac{4 \cdot 10^{-8}[m]}{6 \cdot 10^5 \cdot 10^4[m]}} \cong 1.3 \cdot 10^{-6} cm \equiv 0.013 \mu m \equiv 13 nm \,.$$

This shift corresponds to the complete miss of bunches i.e. mismatch of the order of the beam transverse beam size sigma, so for partial mismatch this number must be reduced at least 10 times for 10% reduction of luminosity, coming to restriction of the order $\Delta y_m \leq 1.3 nm$.

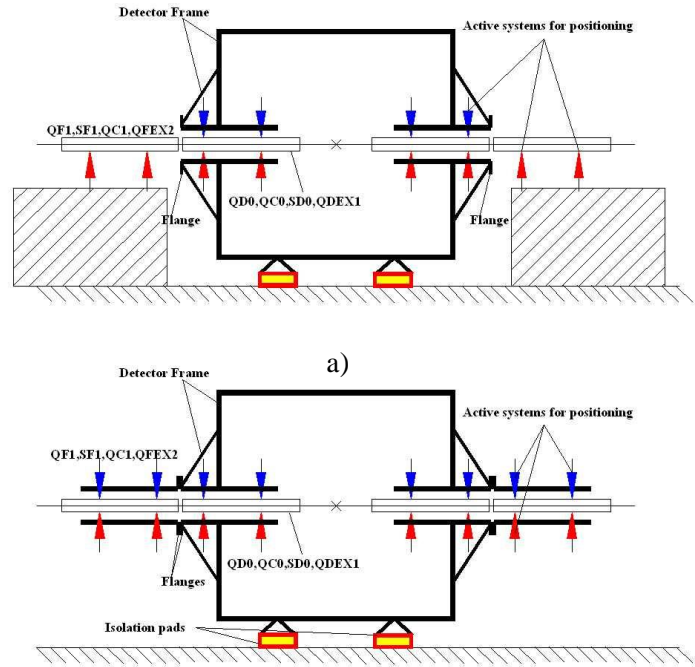

**Figure 17.** a)- Basement of final doublet suggested for ILC, BDR, b)-recommended basement concept.

Indeed, if all lenses participating in beam size formation at both sides of detector move as a whole, this effect does not manifest. That is why we are suggesting installation of all final lenses at the same frame –common practice in ordinary optics (optical table), see Fig.17.

The beam based alignment system, accommodated in ILC will operate a dipole trimming coils mounted inside the same cryostat as the lenses, and will provide equivalent shift of lens axis by changing electrical current in its coils as necessary.

Utilization of 2K Helium in final quads cooling can bring <15% increase of field maximum, so we are not considering it for QD0 at the moment, although it might be introduced later, just widening the margins for safe operation. Other component of the beam optics might include dual bore lenses, if crossing angle at IP required. Example of design of such lens is represented in Fig.18.

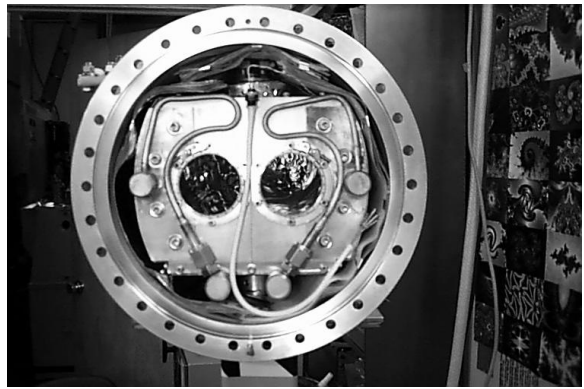

**Figure 18.** Dual bore SC quadrupole developed at Cornell.

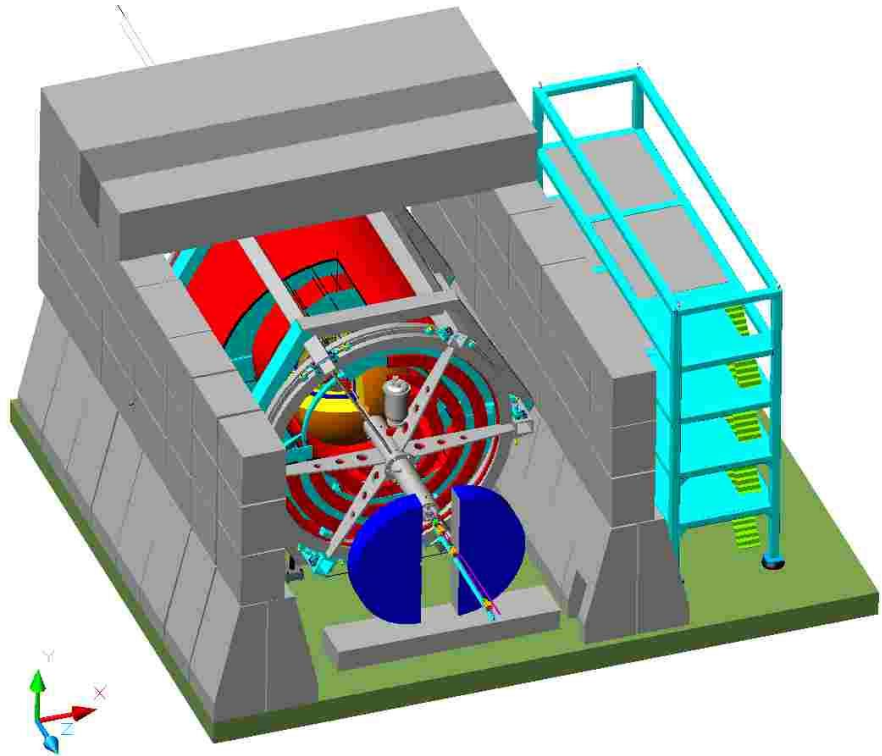

**Figure 19.** Assembled detector inside Borated Concrete walls, made from separate blocks[5].

Important issues associated with Iron-free detector and implemented in a 4$^{th}$ Concept are:
a) Integration of FF hardware into detector;
b) Any crossing angle is OK, but lobby for zero-degree crossing angle with a kicker having travelling wave and BSY for two IRs;
   Easy installation and reinstallations; Numerous experimenta conveniences, e.g., surveying, new add-ons or replacements in later years, etc.
c) Reverse magnetic field in detector to cancel detector asymmetries, especially important for experiments with polarized beams;
d) Possibility of operation with beams $e^+e^-$ or $\mu^+\mu^-$ having different systematic energies of incoming beams (up to few tens $GeV$ difference could be possible).

## *Technology for large solenoids of 4$^{th}$ concept*

As we could see, the return field value depends on the ratio of the areas with corresponding flux. So by making the outer solenoid larger, one can reduce the field, required from outer solenoid and in reaching higher field level in the inner solenoid (Less field value is subtracted).

*Two types* of solenoids design with big diameter are feasible. First is a traditional one with stabilized SC conductor cable (typically with Aluminum), see Fig.21 and Fig.22. The second one is associated with brazing of cable into Aluminum cylinder, see Fig.23.

---

[5] Sometimes existence of few-tenth kiloton Iron yoke exposed as an advantage for shielding of radiation from operating detector.

***Traditional approach*** uses a conductor with dimensions of Al stabilizer ~10x1 *cm*

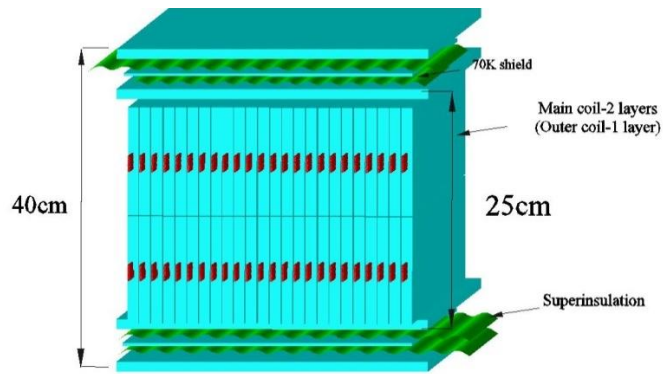

**Figure 20**. Fragment of coil inside a cryostat in a regular part.

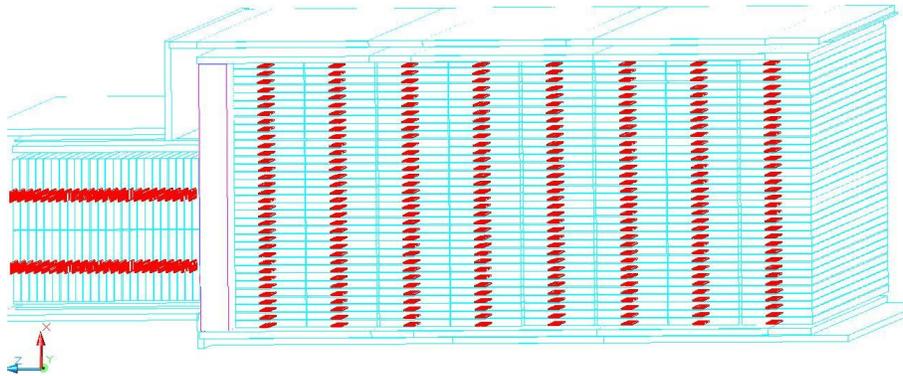

**Figure 21.** End sections of main solenoid made from stabilized conductor. These sections having increased linear current density realizing the Helmholtz coils.

Let us mention also a possibility to split the large-diameter coil in segments, represented in Fig.22.

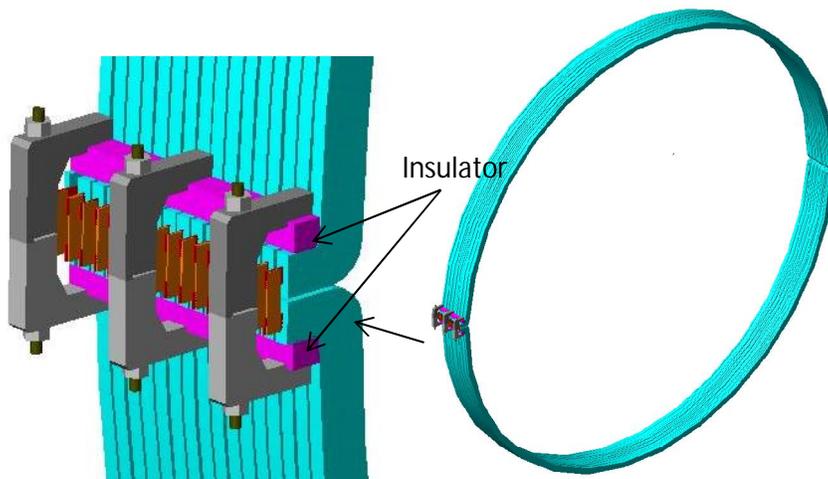

**Figure 22.** The 4$^{th}$ Concept coil with large diameter (outer one) could be split in two (or more) segments.

*Another* approach is to solder the Copper cable directly into the slits in Al cylinder. The solder and flux for brazing Al with Copper is well known [19].

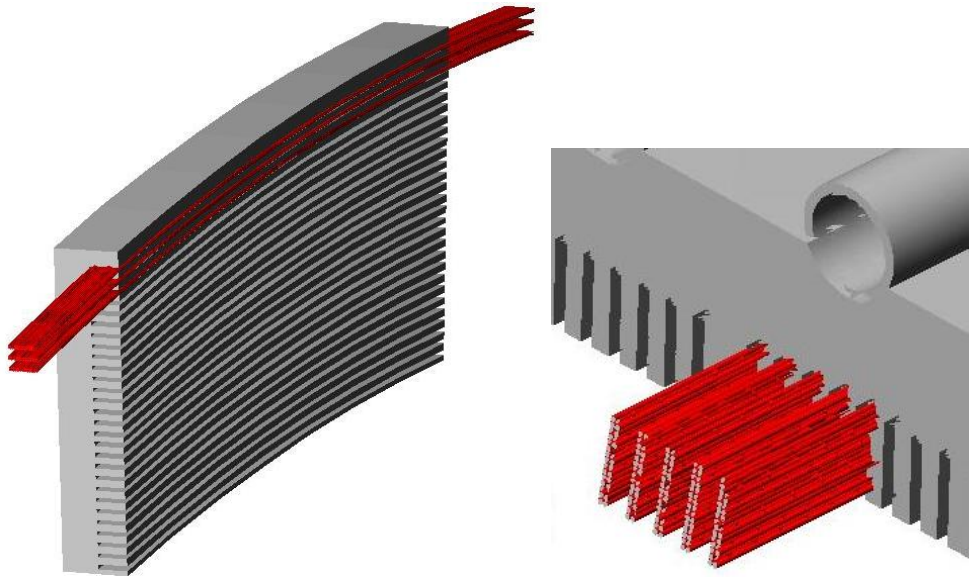

**Figure 23.** The cable soldered in Aluminum carcasses. At the left- geometry for the Helmholtz end coils. At the right-the geometry for regular part of solenoid.

End Helmholtz coils for this type of cabling made as a package of flat discs, where the layering in a disc is going in the radial direction, see Fig.23, left. In a regular part the grooves evenly located along the inner side of cylinder. Such configuration makes positioning of SC cable more rigid. Solenoid could be sectioned in longitudinal (axial) direction inside a common cryostat. Even if all stored energy (~2.8*GJ*) disappeared in carcasses, temperature gain could reach ~70°C, energy evacuation system allows evacuation ~70% of stored energy. Outer solenoid is thinner and has smaller wall-current density; it allows much relaxed design; fragmentation looks feasible, Fig. 22.

For the 4[th] Concept detector the following parameters for solenoid made with brazing technology suggested:

• SC cable with 30=2x15 wires diam. 0.8*mm* each, Total current ~18 *kA* in a cable;
• Cables soldered in tinned grooves made in Al-alloy carcass, Fig.23, right. Separation of grooves is 5*mm* (grooves ~1.5*mm* x 20*mm*), so the spacial period of grooves is 6.5 *mm*;
• Thickness of solenoid (without cryostat) in a regular section ~6*cm*;
• Sectioned assembling; end section ~50*cm* x 13*cm* total in 16 radial sections (Fig.23, left);
• SC cable fixed in grooves with alloy and by compression;
• Indirect cooling of Al carcasses;
• Number of turns in a main solenoid is ~1000x2, the number of turns in outer solenoid is ~500x2

## MULTIPLE FLUX-RETURNS COIL SYSTEM

In this paragraph we will describe the concept of detector with multiple return coils, Fig.2 d). We believe that namely this system with multiple flux-return solenoids is a perspective one, Fig.24 as it allows higher magnetic field on the axis and less expensive solenoid manufacturing and handling.

Formula for the flux redistribution (1) has now a form

$$B_0 \times r_1^2 = N \times B_{rs} \times r_{rs}^2 , \qquad (13)$$

where $B_0$ is a field value in a central solenoid, having radius $r_1$, $B_{rs}$ is a field inside return flux solenoid, having radius $r_{rs}$, N – is a number of return solenoids ($N$=18 in Fig.24).

The outer solenoids could be made segmented-shape cross section, so they will fill practically all cross section, i.e. will be closer to the triple-solenoidal system Fig.2 e) (see also Fig. 33). If the return solenoid has segmented shape with the cross section area $A$, then (13) becomes

$$B_0 \times r_1^2 = \tfrac{1}{\pi} N \times B_{rs} \times A \qquad (13a)$$

Inside small return-flux solenoids the drift chamber systems for muon spectrometry are installed. They serve for spectrometry of muons.

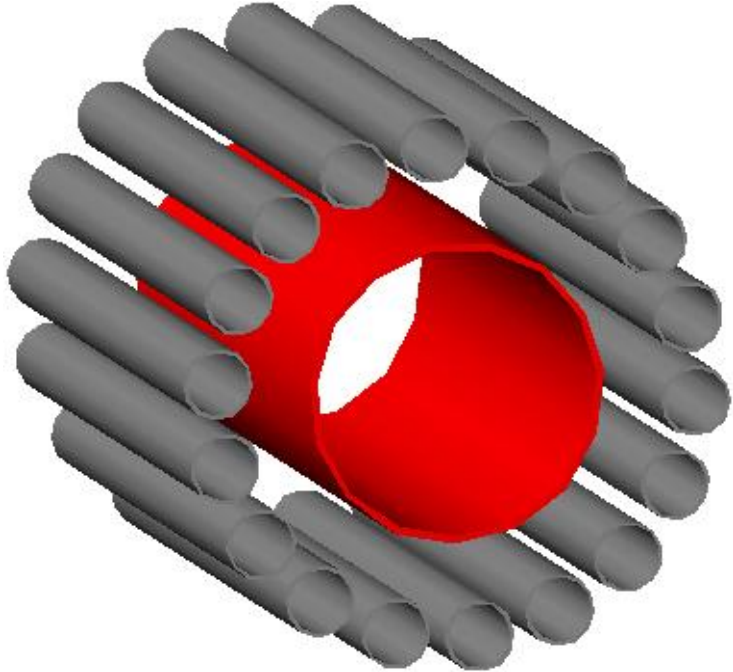

**Figure 24.** This is a 3D isometric view on the magnetic system with multiple flux-return solenoids. Central solenoid painted in red.

The spacing between main solenoid and multiple flux-return ones defined by technological requirements, although some tracking system could be located there also (no-magnetic field region). Distribution of currents is represented in Fig.25 and Fig.26.

We calculated magnetic field distribution in a simplest system with solenoids having round cross section for simplicity; any other shape will give the same principal results. In principle these flux-return solenoids could be arranged as a honeycomb-like structure.

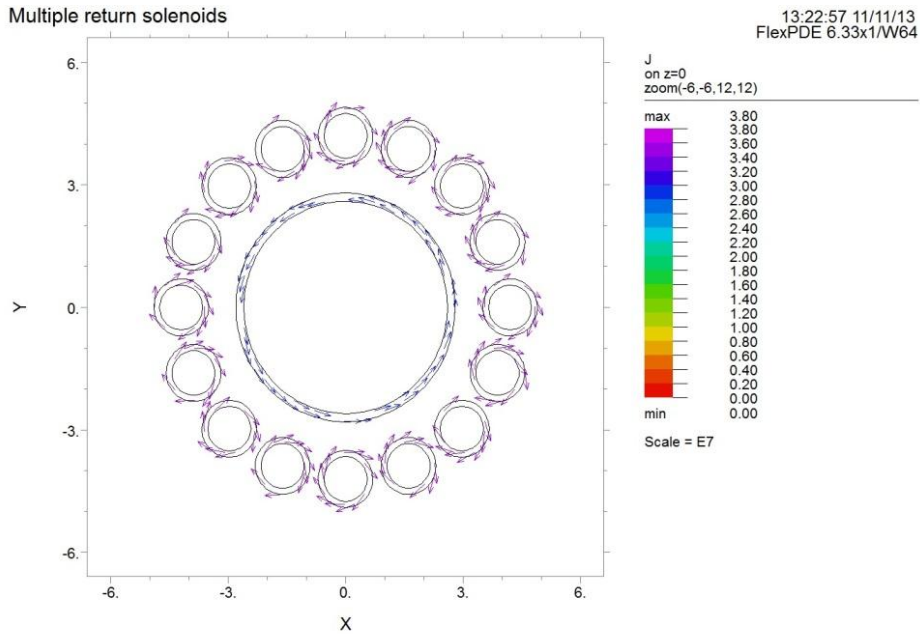

**Figure 25.** Distribution of currents shown by vectors, top view.

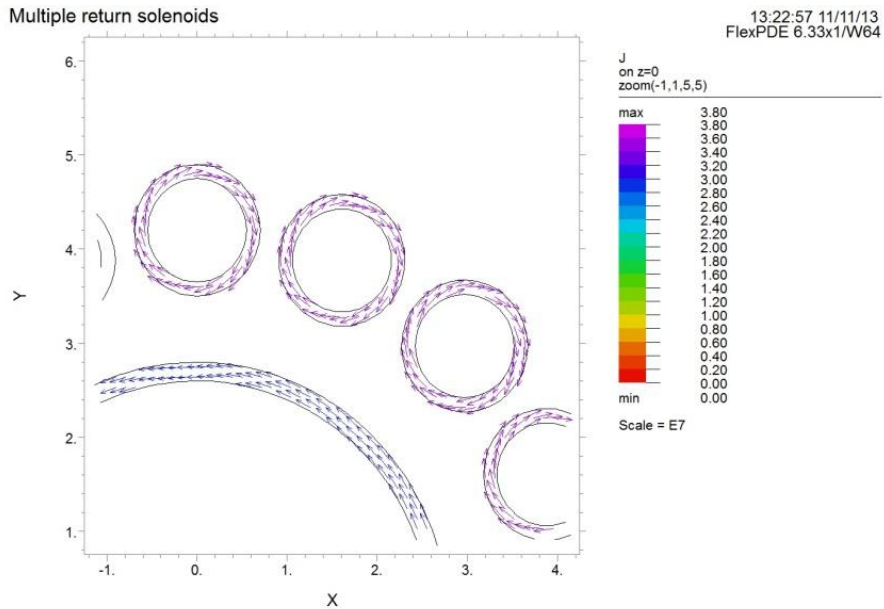

**Figure 26.** Distribution of currents, Fig.25 zoomed.

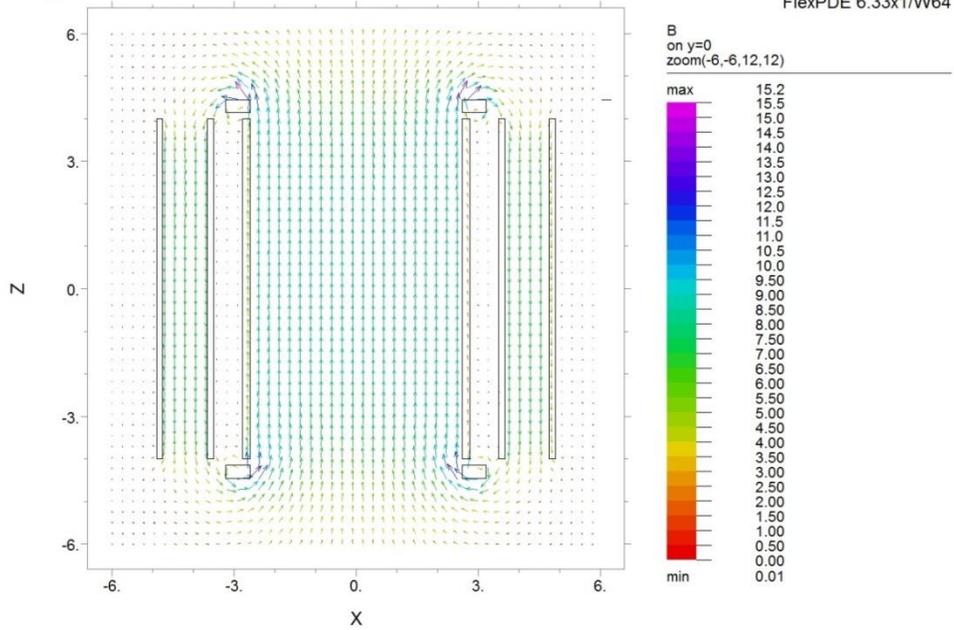

**Figure 27.** Vectors of magnetic field.

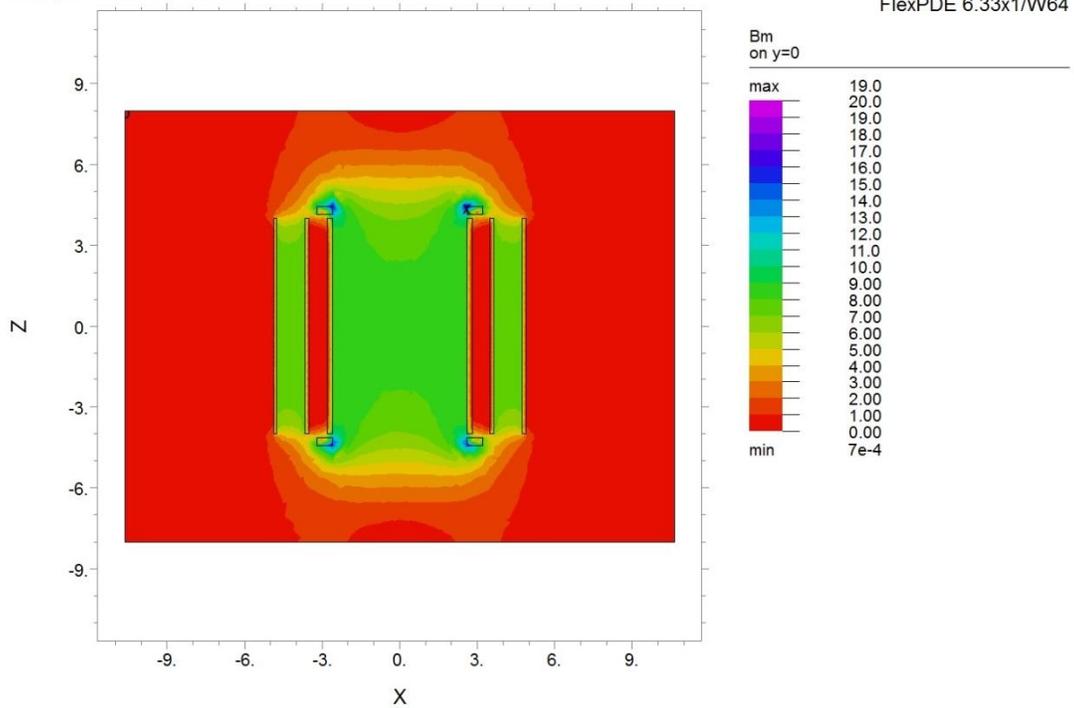

**Figure 28.** Magnetic field amplitude contour, painted.

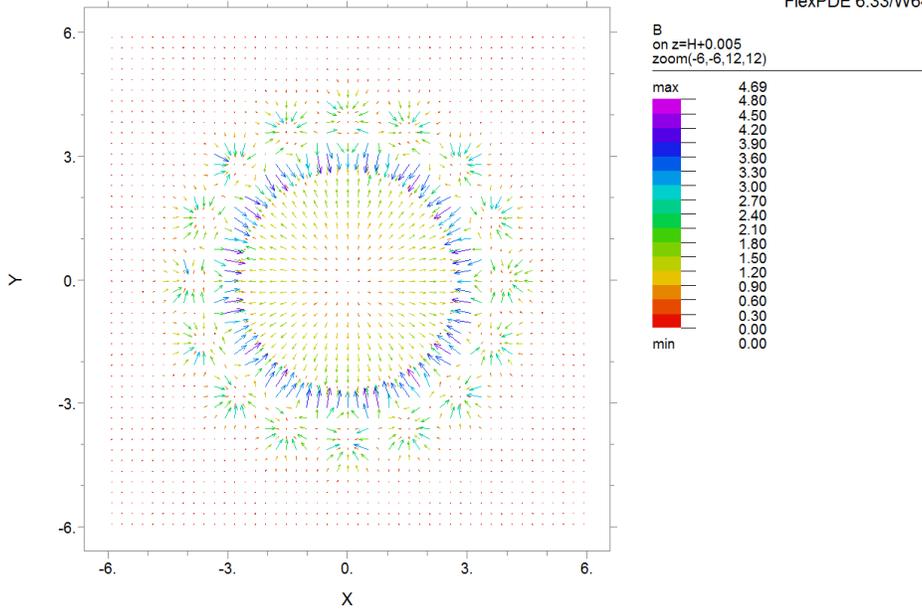

**Figure 29.** Vectors of magnetic field in a plane just above cutoff return-flux solenoids.

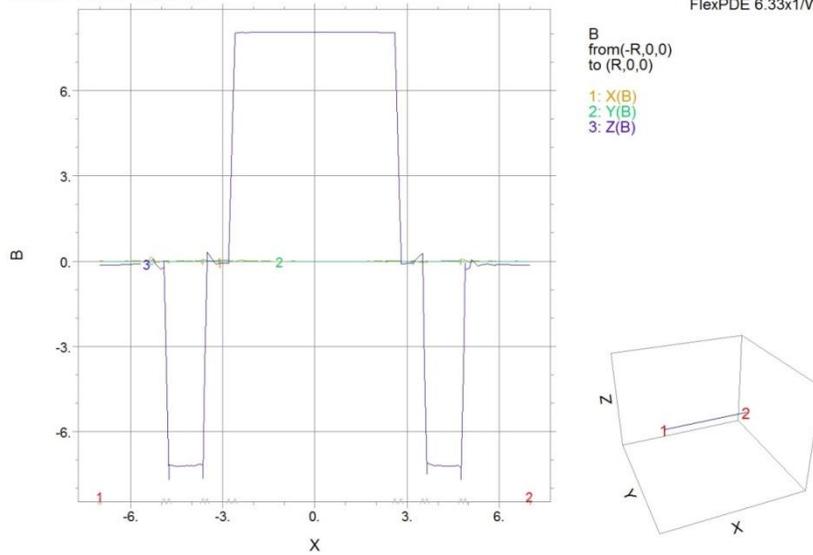

**Fgure 30**. Elevation of magnetic field across in a midplane (FlexPDE).

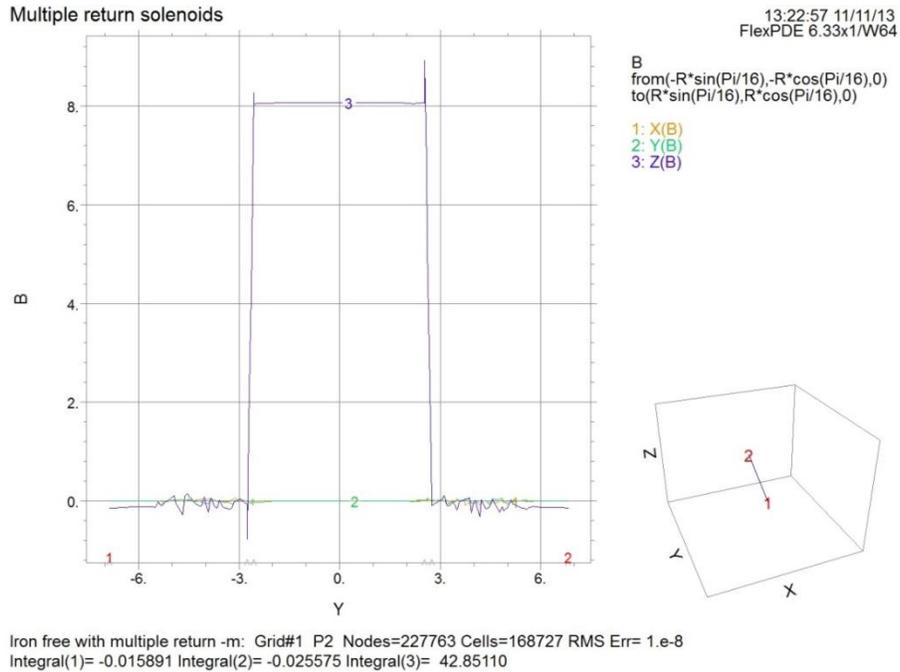

**Figure 31.** Field across, but now the elevation plotted for the line which runs between the return-flux solenoids.

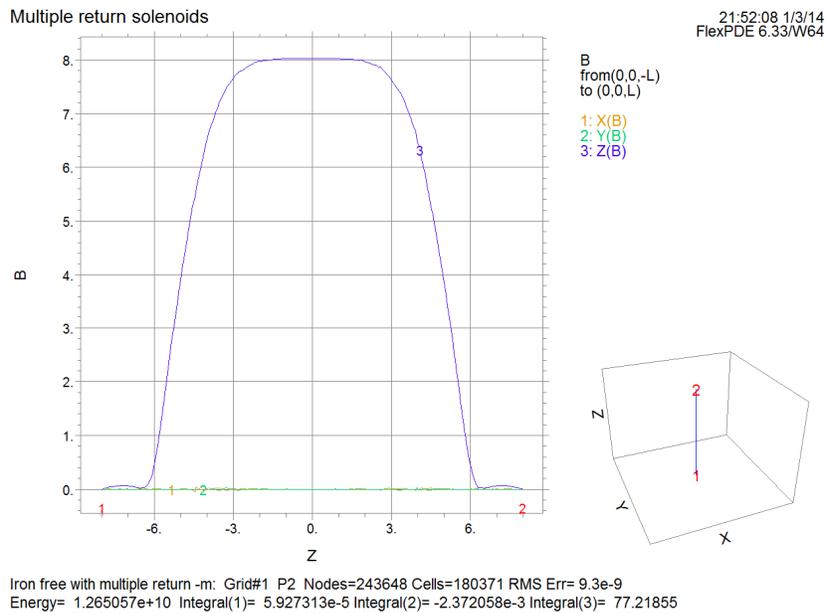

**Figure 32**. Elevation of magnetic field along central axis. Field value in *Tesla*.

As we mentioned, the outer solenoids could be made segmented, so they will fill practically all volume, i.e. will be closer to the triple solenoidal system, Fig.33, Fig. 2-e). The loss of solid angle could be made <10%. Technology of winding is practically the same as for solenoids with cylindrical shape.

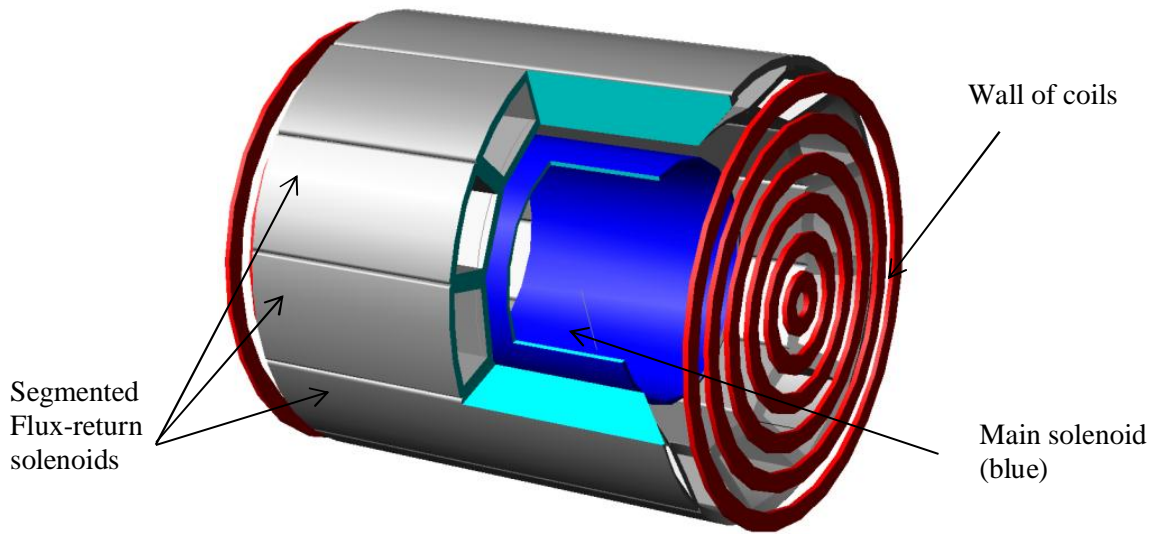

**Figure 33.** Many return-flux solenoids with the shape of segments. This is done for better coverage the volume with magnetic field.

Magnetic field with total current in central solenoid ~the same current density as in $4^{th}$, now comes to be ~$8T$.

Immersed into the frame, supporting the wall of coils also, this detector looks pretty much the same as $4^{th}$ Concept detector is, see Fig.34

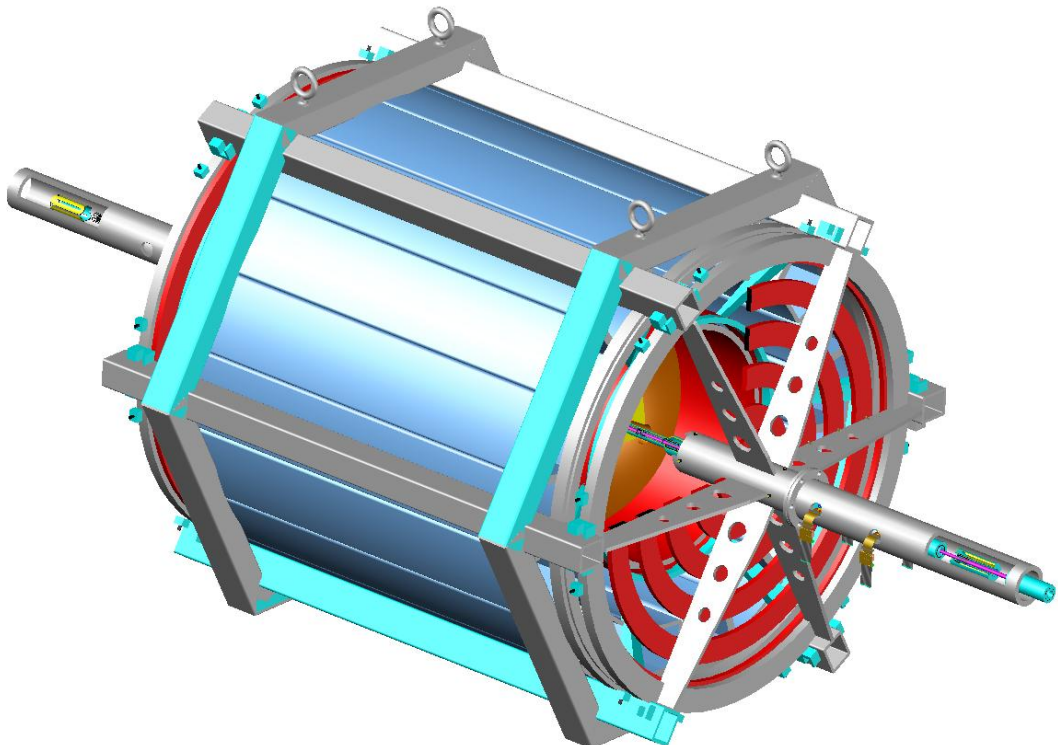

**Figure 34.** Iron free detector with multiple flux-return solenoids.

Final lenses are carried by the same reinforced frame. This detector can accommodate the laser driven collider easily, see [18].

## CONCLUSIONS

Detector system with multiple flux-return solenoids allows

1) Lightweight configuration with easily accessed structure, modular design, mobile.
2) It is less expensive, than 4$^{th}$ Concept and traditional detectors with Iron yoke.
3) Flexibility for rearrangement of inner configuration for serving specitic requirements of experiment (asymmetries while operating with polarized particles, different energies of colliding beams for better spatial resolution of collision vertexes, etc.)
4) Higher magnetic field at IP, as there is no subtraction of field by outer solenoid.

Absence of saturated thick Iron yoke makes possible pretty quick reversing of field in a detector; this might be useful for exclusion of asymmetries of registration system.

Possible commercial usage of such system (for MRI) opens a possibility to combine efforts at the stage of magnet design.

All technologies applicable for 4$^{th}$ could be implemented into system with multiple-flux return solenoidal system.

Of cause, in a brief report we illuminated only small fraction of interesting possibilities opened by a multiple flux-return solenoidal magnet system. The other systems (calorimeters, tracking, etc.) require a lot of professional attention.